\documentclass[preprintnumbers,amsmath,amssymbm,prd]{revtex4}
\usepackage{epsfig}
\usepackage{graphicx}
\usepackage{amssymb}

\begin{document}
\title{Black-hole perturbation theory: The asymptotic spectrum of the prolate spin-weighted spheroidal
harmonics}
\author{Shahar Hod}
\address{The Ruppin Academic Center, Emeq Hefer 40250, Israel}
\address{ }
\address{The Hadassah Institute, Jerusalem 91010, Israel}
\date{\today}

\begin{abstract}
\ \ \ Prolate spin-weighted spheroidal harmonics play a key role in
black-hole perturbation theory. In particular, the highly damped
quasinormal resonances of rotating Kerr black holes are closely
related to the asymptotic eigenvalues of these important functions.
We here present a novel and compact derivation of the asymptotic
eigenvalues of the prolate spin-weighted spheroidal harmonics. Our
analysis is based on a simple trick which transforms the
corresponding spin-weighted spheroidal angular equation into a
Schr\"odinger-like wave equation which is amenable to a standard WKB
analysis. Our {\it analytical} results for the prolate asymptotic
spectrum agree with previous {\it numerical} computations of the
eigenvalues which appear in the literature.
\end{abstract}
\bigskip
\maketitle

%]

\section{Introduction.}

The characteristic dynamics of test fields in black-hole spacetimes
has been studied extensively since the pioneering work of Regge and
Wheeler \cite{ReWh}, see also \cite{Nollert1,Press,Tails} and
references therein. An astrophysically realistic model of wave
dynamics in black-hole spacetimes should involve a non-spherical
background geometry with angular momentum. In terms of the
Boyer-Lindquist coordinates, the spacetime of a rotating Kerr black
hole is described by the line-element \cite{Chan,Kerr}
\begin{eqnarray}\label{Eq1}
ds^2=-\Big(1-{{2Mr}\over{\rho^2}}\Big)dt^2-{{4Mar\sin^2\theta}\over{\rho^2}}dt
d\phi+{{\rho^2}\over{\Delta}}dr^2
%\nonumber \\
+\rho^2d\theta^2+\Big(r^2+a^2+{{2Ma^2r\sin^2\theta}\over{\rho^2}}\Big)\sin^2\theta
d\phi^2,
\end{eqnarray}
where $M$ and $a$ are the mass and angular momentum per unit mass of
the black hole, respectively. (We use gravitational units in which
$G=c=1$). Here $\Delta\equiv r^2-2Mr+a^2$ and $\rho\equiv
r^2+a^2\cos^2\theta$.

In this paper we consider perturbations of the {\it non}-spherical
Kerr spacetime. The dynamics of a test field $\Psi$ in the rotating
Kerr spacetime is governed by the well-known Teukolsky master
equation \cite{Teu}. One may decompose the field as
\begin{equation}\label{Eq2}
_s\Psi_{lm}(t,r,\theta,\phi)=e^{im\phi}{_sS_{lm}}(\theta;a\omega){_s\psi_{lm}}(r)e^{-i\omega
t}\ ,
\end{equation}
where $\omega$ is the (conserved) frequency of the mode, $l$ is the
spheroidal harmonic index, and $m$ is the azimuthal harmonic index.
The parameter $s$ is called the spin weight of the field, and is
given by $s=\pm 2$ for gravitational perturbations, $s=\pm 1$ for
electromagnetic perturbations, $s=\pm {1\over 2}$ for massless
neutrino perturbations, and $s=0$ for scalar perturbations
\cite{Teu}.

With the decomposition (\ref{Eq2}), $\psi$ and $S$ obey radial and
angular equations, both of confluent Heun type
\cite{Teu,Heun,Fiz1,Abram,Flam}, coupled by a separation constant
$A(a\omega)$. The radial Teukolsky equation is given by \cite{Teu}
\begin{equation}\label{Eq3}
\Delta^{-s}{{d}
\over{dr}}\Big(\Delta^{s+1}{{d\psi}\over{dr}}\Big)+\Big[{{K^2-2is(r-M)K}\over{\Delta}}
-a^2\omega^2+2ma\omega-A+4is\omega r\Big]\psi=0\ ,
\end{equation}
where $K\equiv (r^2+a^2)\omega-am$.

The angular functions $S(\theta;a\omega)$ are the spin-weighted
spheroidal harmonics which are solutions of the angular equation
\cite{Teu,Heun,Fiz1,Abram,Flam}
\begin{equation}\label{Eq4}
{1\over {\sin\theta}}{\partial \over
{\partial\theta}}\Big(\sin\theta {{\partial
S}\over{\partial\theta}}\Big)+\Big[c^2\cos^2\theta-2c
s\cos\theta-{{(m+s\cos\theta)^2}\over{\sin^2\theta}}+s+A\Big]S=0\  ,
\end{equation}
where $c\equiv a\omega$.
%\in\mathbb{C}$
The angular functions are
required to be regular at the poles $\theta=0$ and $\theta=\pi$.
These boundary conditions pick out a discrete set of eigenvalues
$\{_sA_{lm}\}$ labeled by the integers $m$ and $l$. [In the $c\to 0$
limit these angular functions become the familiar spin-weighted
spherical harmonics with the corresponding angular eigenvalues
$A=l(l+1)-s(s+1)+O(a\omega)$.]

The spin-weighted spheroidal harmonics $S(\theta;c)$ and their
corresponding eigenvalues $\{_sA_{lm}\}$ have attracted much
attention over the years from both physicists and mathematicians
\cite{Flam,Meix,Breu1,Breu2,Cas,BCC,Hodo}. It is worth emphasizing
that in order to compute the characteristic resonances of black
holes \cite{Nollert1}, one must {\it first} compute the (closely
related) angular eigenvalues $\{_sA_{lm}\}$ [see Eq. (\ref{Eq3})].
In particular, in the framework of semi-classical general
relativity, it has been conjectured that the highly damped
resonances may shed light on the quantum properties of black holes
\cite{HodL1,HodL2}. For rotating black holes, these highly damped
resonances are characterized by $c_I\to\infty$ (where $c_I\equiv \Im
c$).

The asymptotic limit $c_I\to\infty$ of the angular equation
(\ref{Eq4}) was studied in \cite{Flam,Meix} for the $s=0$ case. It
was found that the asymptotic scalar eigenvalues are given by
\begin{equation}\label{Eq5}
_0A_{lm}=[2(l-|m|)+1] |c_I|+O(1)\  .
\end{equation}
As pointed out in \cite{BCC}, the analysis of \cite{Flam,Meix} for
the spin-$0$ case is somewhat incomplete. The analysis of
\cite{Flam,Meix} requires the knowledge of the number of zeros of
the scalar harmonics in the interval $[0,\pi]$. However, as
emphasized in \cite{BCC}, the approximated solution found in
\cite{Flam,Meix} for the $c_I\to\infty$ limit is only valid in a
region far from the end-points $\theta=0,\pi$. Thus, the analysis of
\cite{Flam,Meix} cannot rule out the possible existence of
additional zeros of the angular eigenfunctions near the end-points.
The possible omission of such zeros would lead to a wrong asymptotic
behavior of the prolate (with $c=ic_I$) eigenvalues, see \cite{BCC}
for details. In this respect the analysis of \cite{Flam,Meix} is not
complete. One of the goals of the present paper is to present a more
rigorous proof of the formula (\ref{Eq5}) for the asymptotic scalar
eigenvalues.
%Moreover, we
%believe that our proof is also simpler and more intuitive than the
%one given in \cite{Flam,Meix} for the spin-$0$ case.

The asymptotic spectrum of the prolate eigenvalues for the general
spin case was first studied in \cite{Breu1}. However, as pointed out
in \cite{BCC,BerCarY}, the analysis of \cite{Breu1} for the general
spin case is fundamentally flawed (see \cite{BCC,BerCarY} for
details). The {\it numerical} results presented in Refs.
\cite{BCC,BerCarY} for the general spin case provide evidence for
the asymptotic behavior
\begin{equation}\label{Eq6}
_sA_{lm}=\gamma |c_I|+O(1)\  ,
\end{equation}
where the function $\gamma$ depends on the spin-parameter $s$, the
azimuthal harmonic index $m$, and the spheroidal harmonic index $l$
[see Eq. (\ref{Eq21}) below]. The main aim of the present paper is
to provide a (simple) {\it analytical} proof for the prolate formula
(\ref{Eq6}) in the {\it general} spin case.

\section{A coordinate transformation}

It proves useful to introduce the coordinate $x$ defined by
\cite{Yang,Hodo}
\begin{equation}\label{Eq7}
x\equiv\ln\Big(\tan\Big({{\theta}\over{2}}\Big)\Big)\  ,
\end{equation}
in terms of which the angular equation (\ref{Eq4}) becomes a
Schr\"odinger-like wave equation of the form \cite{NoteSch}
\begin{equation}\label{Eq8}
{{d^2S}\over{dx^2}}-US=0\  ,
\end{equation}
where
\begin{equation}\label{Eq9}
U(\theta)=(m+s\cos\theta)^2-\sin^2\theta(c^2\cos^2\theta-2cs\cos\theta+s+A)\
.
\end{equation}
Note that the interval $\theta\in [0,\pi]$ maps into
$x\in[-\infty,\infty]$. The Schr\"odinger-type angular equation
(\ref{Eq8}) is now in a form that is amenable to a standard WKB
analysis.

\section{The spin-$0$ (scalar) case.}

We shall first consider the spin-0 (scalar) case. In this case the
effective potential $U(\theta)$ is in the form of a symmetric
(invariant under the transformation $\theta\to\pi-\theta$) potential
well: in the $c_I\to\infty$ limit it has a local minimum at
\begin{equation}\label{Eq10}
{_0\theta_{\text{min}}}={\pi\over 2}\ \ \ \text{with} \ \ \
U({_0\theta_{\text{min}}})=-A+O(1)\  .
\end{equation}

Regions where $U(\theta)<0$ are characterized by an oscillatory
behavior of the wave-function $S$ (the `classically allowed
regions'), while regions with $U(\theta)>0$ (the `classically
forbidden regions') are characterized by an exponential behavior
(evanescent waves). The `classical turning points' are characterized
by $U=0$. There is a pair $\{_0\theta^-,_0\theta^+\}$ of such
turning points (with
${_0\theta^-}<{_0\theta_{\text{min}}}<{_0\theta^+}$), which in the
$c_I\to\infty$ limit are located in the near vicinity of
$_0\theta_{\text{min}}$:
\begin{eqnarray}\label{Eq11}
{_0\theta^{\pm}}={\pi\over 2} \pm
{{\sqrt{A}}\over{|c_I|}}+O(c^{-3/2}_I)\ .
\end{eqnarray}

A standard textbook second-order WKB approximation for the
bound-state `energies' of a Schr\"odinger-like wave equation of the
form (\ref{Eq8}) yields the well-known quantization condition
\cite{WKB1,WKB2,WKB3,Iyer,Notehigh}
\begin{equation}\label{Eq12}
\int_{x^{-}}^{x^{+}}dx\sqrt{-U(x)}=(N+{1\over 2})\pi\ \ \ ; \ \ \
N=0,1,2,...\  ,
\end{equation}
where $x^{\pm}$ are the turning points [with $U(x^{\pm})=0$] of the
potential well, and $N$ is a non-negative integer.

Using the relation $dx/d\theta=1/\sin\theta$, one can write the WKB
condition (\ref{Eq12}) in the form
\begin{equation}\label{Eq13}
\int_{{_0\theta^{-}}}^{{_0\theta^{+}}}
d\theta{{\sqrt{-U(\theta)}}\over{\sin\theta}}=(N+{1\over
2})\pi\ \ \ ; \ \ \ N=0,1,2,...\  .
\end{equation}
The WKB quantization condition (\ref{Eq13}) determines the
eigenvalues $\{A\}$ of the associated spin-weighted spheroidal
harmonics in the large-$c_I$ limit. The relation so obtained between
the eigenvalues and the parameters $c,m,s$ and $N$ is rather complex
and involves elliptic integrals. However, given the fact that in the
$c_I\to\infty$ limit the turning points $_0\theta^{\pm}$ lie in the
vicinity of $\theta={\pi\over 2}$ [see Eq. (\ref{Eq11})], one can
approximate the integral in (\ref{Eq11}) by \cite{Noteapp}
\begin{eqnarray}\label{Eq14}
\int_{{_0\theta^{-}}}^{{_0\theta^{+}}}d\theta
\sqrt{-c^2_I(\theta-{\pi\over 2})^2+A}=(N+{1\over 2})\pi\ \ \ ; \ \
\ N=0,1,2,...\  .
\end{eqnarray}
Evaluating the integral in (\ref{Eq14}) is straightforward, and one
finds
\begin{equation}\label{Eq15}
A(N)=(2N+1)|c_I|+O(1)\ \ \ ; \ \ \ N=0,1,2,...\
\end{equation}
for the quantized spectrum. This completes our proof for the prolate
asymptotic spectrum in the scalar ($s=0$) case \cite{NoteCas}.

\section{The general spin case.}

We shall now consider the general spin case. In this case the
effective potential $U(\theta)$ is complex-valued. Its minimum is
located at $_s\theta_{\text{min}}={\pi\over
2}+{{is}\over{c_I}}+O(c^{-2}_I)$, while the two turning points are
located at
\begin{eqnarray}\label{Eq16}
_s\theta^{\pm}={\pi\over 2} \pm
{{\sqrt{A}}\over{|c_I|}}+{{is}\over{c_I}}+O(c^{-3/2}_I)\ .
\end{eqnarray}

A natural generalization of the WKB analysis to the case of
complex-valued potentials is provided in \cite{Miller}: the WKB
quantization rule is given by the standard relation
\begin{equation}\label{Eq17}
\int_{_s\theta^{-}}^{_s\theta^{+}}d\theta{{\sqrt{-U(\theta)}}\over{\sin\theta}}=(N+{1\over
2})\pi\ \ \ ; \ \ \ N=0,1,2,...\  ,
\end{equation}
which can be approximated near $\theta={\pi/2}$ by \cite{Noteapp}
\begin{eqnarray}\label{Eq18}
\int_{_s\theta^{-}}^{_s\theta^{+}}d\theta
\sqrt{-c^2_I(\theta-{\pi\over 2})^2-2isc_I(\theta-{\pi\over
2})+A}=(N+{1\over 2})\pi\ \ \ ; \ \ \ N=0,1,2,...\  .
\end{eqnarray}
As emphasized in \cite{Miller}, the integration path between the two
complex turning points $\{_s\theta^-,_s\theta^+\}$ should be chosen
such that
\begin{equation}\label{Eq19}
\Im\{{\sqrt{-U(\theta)}}\}=0
\end{equation}
along the integration contour \cite{Miller}. In the $c_I\to\infty$
limit this requirement is easily fulfilled by a straight line
(parallel to the real $\theta$-axis) which connects the two turning
points (\ref{Eq16}). Substituting $\theta=\phi-{{is}\over{c_I}}$
(where $\phi\in\mathbb{R}$ runs from ${_0\theta^-}$ to
${_0\theta^+}$) into (\ref{Eq18}) and neglecting terms of order
$O(1)$, one finds that along the path (\ref{Eq19}) the integral
(\ref{Eq16}) can be written as
\begin{eqnarray}\label{Eq20}
\int_{{_0\theta^{-}}}^{{_0\theta^{+}}}d\phi
\sqrt{-c^2_I(\phi-{\pi\over 2})^2+A}=(N+{1\over 2})\pi\ \ \ ; \ \ \
N=0,1,2,...\  .
\end{eqnarray}
This yields
\begin{equation}\label{Eq21}
A(N)=(2N+1)|c_I|+O(1)\ \ \ ; \ \ \ N=0,1,2,...\
\end{equation}
for the quantized spectrum. This completes our proof for the prolate
asymptotic spectrum in the general spin case \cite{NoteCas}. It is
worth emphasizing that the {\it analytical} formula (\ref{Eq21}) for
the prolate asymptotic spectrum agrees with the {\it numerical}
results presented in \cite{BCC,BerCarY}.

%agree with the analytical results of \cite{Flam,Meix} for the
%specific case of spin-$0$ (scalar) fields.

\bigskip
\noindent
{\bf ACKNOWLEDGMENTS}
\bigskip

This research is supported by the Carmel Science Foundation. I thank
Uri Keshet, Oded Hod, Yael Oren, Arbel M. Ongo and Ayelet B. Lata
for helpful discussions.


\begin{thebibliography}{99}
\bibitem{NoteSch} Note that in the quantum-mechanical terminology
$-U$ stands for ${{2m}\over{\hbar^2}}(E-V)$, where $E, V$, and $m$
are the total energy, potential energy, and mass of the of the
particle, respectively.
\bibitem{Notehigh} Higher order corrections to the asymptotic eigenvalues [see formula (\ref{Eq18}) below] can
be obtained by using a higher-order WKB analysis \cite{Iyer}.
\bibitem{Noteapp} Here we have used the relations
$\sin\theta=1-{1\over 2}(\theta-{\pi\over 2})^2+O[(\theta-{\pi\over
2})^4]$ and $\cos\theta=-(\theta-{\pi\over 2})+O[(\theta-{\pi\over
2})^3]$.
\bibitem{NoteCas} Note that our discrete resonance parameter $N$ corresponds
to $l-\text{max}(|m|,|s|)$ of \cite{Flam,Meix}.
\end{thebibliography}
\end{document}